# Spin dephasing in Silicon Germanium ($Si_{1-x}Ge_x$) nanowires

Ashish Kumar[a)] and Bahniman Ghosh

*Department of Electrical Engineering, Indian Institute of Technology, Kanpur 208016, India*

Abstract – We study spin polarized transport in silicon germanium nanowires using a semiclassical monte carlo approach. Spin depolarization in the channel is caused due to D'yakonov-Perel (DP) relaxation associated with Rashba spin orbit coupling and due to Elliott-Yafet (EY) relaxation. We investigate the dependence of spin dephasing on germanium mole fraction in silicon germanium nanowires. The spin dephasing lengths decrease with an increase in the germanium mole fraction. We also find that the temperature has a strong influence on the dephasing rate and spin relaxation lengths increase with decrease in temperature. The ensemble averaged spin components and the steady state distribution of spin components vary with initial polarization.

Keywords: Spin Dephasing, Monte Carlo, Silicon Germanium Nanowires, Mole fraction

[a] Electronic) mail : ashish12.kumar@gmail.com
  Phone No.: +91-9415858224

I. INTRODUCTION

There has been recently growing interest in the field of spintronics that makes use of the spin degree of freedom compared to the conventional usage of the charge degree of freedom [1-4]. A number of spin based devices [5-9] have also been proposed that use this concept to realize conventional electronic devices. The considerable interest in this field is because of the advantages that are inherent with the usage of this quantum property of an electron. Operation of spintronic deices require lesser power and dissipate lesser power as compared to charge based electronic devices. Using spin to store information also leads to faster processing speeds. Moreover spin lends itself beautifully for quantum computation [10-12].

Any spintronic device fundamentally operates on three basic processes- spin injection, spin transport and spin detection. In our work we focus on spin transport in semiconductors. Determination of spin transport properties of a material constitutes a critical part for establishing the suitability of a semiconductor for transmitting information in spin based devices. Spin relaxation study forms an integral part of determining spin transport properties. Researchers have already established that spin relaxation times in semiconductors are much larger [13-14] compared to momentum (charge) relaxation times and is another motivation strong enough to propel our move towards spintronics.

In this paper we study spin polarized transport in nanowires (one dimensional structure) using semiclassical Monte Carlo approach. In the past also great deal of theoretical and experimental work [15-24] has been done to ascertain spin transport properties in materials of potential interest. Recently we conducted our work to determine spin relaxation lengths in silicon [25] using semiclassical monte carlo simulations and our continuing work on other materials of potential interest. However to the best of our knowledge, there is no work at present that deals with tackling the issue of spin dephasing in silicon germanium alloy which has of late surged to be a potent electronic material. Silicon Germanium technology has emerged as a viable technology over the Si technology. Silicon is a very useful material from the integrated circuit manufacturing point of view because of the fact that Si has a high quality dielectric ($SiO_2$) which can be grown very easily on it. However silicon has low carrier mobility and a low saturation velocity which limit the maximum frequency of operation. This can be get rid of by alloying Si with Ge enables us to achieve a higher carrier mobility and higher saturation mobility than Si. By varying the Ge mole fraction in the alloy the bandgap can be very easily reduced as compared to Silicon and thus can provide many advantages over Si technology. Thus SiGe technology helps to maintain the fabrication advantages of silicon and at the same time gives improved performance over Si.

Over the years many quantum mechanical [9,27] and classical models (drift diffusion models) [28-29] have been proposed to study spin transport. In our work we make use of the semiclassical Monte Carlo [30-31] approach since the evolution of spin is coupled with the evolution of electron momentum. This can best be treated by using a Monte Carlo simulation which has widely been used to model electron transport in devices. The evolution of momentum is treated using a standard Monte Carlo approach which can then be used along with spin matrix calculations for spin evolution [15,17,20,21,23,24].

In this paper we investigate spin dephasing in SiGe. We undertake a study on the dependence of spin dephasing lengths on Ge mole fraction. We also investigate the dependence on temperature. The variation of spin components for different initial polarization is also studied in our work. The paper is organized as follows. In the next section we discuss the model used for simulation in our work. In section III we present our results along with the discussion on them. Finally in section IV we present our conclusion.

II. MODEL

A comprehensive account of the Monte Carlo simulations [24,30,31] and spin transport model [15,20,24] is described elsewhere. In this paper we shall restrict ourselves to discussing only the necessary features of the model and the key modifications. Fig.1 shows the nanowire structure and the co-ordinate system chosen.

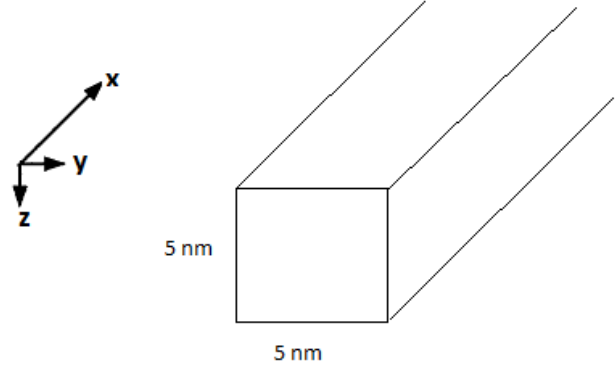

Fig.1 Geometry of the nanowire and the co-ordinate axes.

Both silicon and germanium are elemental semiconductors and hence they possess bulk inversion symmetry [19,22]. Thus the spin orbit coupling due to bulk inversion asymmetry (Dresselhaus spin orbit interaction) is absent [19,22] in both and as a result in the silicon germanium alloy. However the transverse electric field breaks the structural inversion symmetry and leads to spin orbit coupling due to structural inversion asymmetry (Rashba spin orbit interaction). This Rashba spin orbit interaction results in D'yakonov-Perel (DP) relaxation[32] and leads to spin depolarization along the channel. We are interested in how the injected spin polarization decays along the channel under the influence of the driving electric field along the channel length and transverse electric field perpendicular to the channel length.

The temporal evolution of the spin vector is governed by the following equation [20] which governs the spin dynamics during the free flight time,

$$\frac{d\vec{S}}{dt} = \vec{\Omega} \times \vec{S} \qquad (1)$$

where the precession vector $\vec{\Omega}$ has contribution only due to the Rashba Hamiltonian and can be written as [25],

$$\Omega_R(k_x)^{1D} = -\frac{2\alpha k_x \hat{j}}{\hbar} \qquad (2)$$

where the Rashba coefficient $\alpha$ is given by [33]

$$\alpha = \frac{\hbar^2}{2m^*} \frac{\Delta}{E_g} \frac{2E_g+\Delta}{(E_g+\Delta)(3E_g+2\Delta)} eE \qquad (3)$$

where $\Delta$ is the spin orbit splitting of the valence band, $e$ is the electronic charge, $m^*$ is the effective mass, $E_g$ is the band gap and $E$ is the transverse electric field. Here the dependence of spin orbit splitting and band gap on Ge mole fraction is taken into account.

Using Eq. (2) in Eq. (1) and expressing spin vector $\vec{S} = S_x \hat{\imath} + S_y \hat{\jmath} + S_z \hat{k}$ we arrive at following relations

$$\frac{dS_x}{dt} = -\frac{2}{\hbar} \alpha k_x S_z \qquad (4)$$

$$\frac{dS_y}{dt} = 0 \qquad (5)$$

$$\frac{dS_z}{dt} = \frac{2}{\hbar} \alpha k_x S_x \qquad (6)$$

Spin dephasing also occurs because of Elliott-Yafet spin relaxation [34]. EY relaxation is included as an instantaneous spin flip scattering in our simulation and the relaxation time for EY mechanism is taken from Ref. [35]

$$\frac{1}{\tau_s^{EY}} = A \left(\frac{k_B T}{E_g}\right)^2 \eta^2 \left(\frac{1-\eta/2}{1-\eta/3}\right)^2 \frac{1}{\tau_p} \qquad (7)$$

where $\eta = \Delta/(E_g + \Delta)$ and $\tau_p$ being the momentum relaxation time.

For *x*<0.85 the conduction band structure resembles that of silicon and the conduction band minimum lies in the (100) direction or X-valleys [26]. Thus we make use of the silicon band structure for Monte Carlo simulations and as explained in [25], the conduction band splits and we consider only the lower 4 valleys for the sake of our simulation

The scattering processes considered are intravalley and intervalley phonon scattering, surface roughness scattering, ionized impurity scattering and alloy scattering. Both optical phonons and acoustic phonons have been considered. The formula for computation of scattering rates are taken from references [36,37].We use both Si and Ge phonon modes by weighing the corresponding phonon scattering rates by *(1-x)* and *x* [38]

The alloy scattering rate when an electron scatters from the $n^{th}$ subband to the $m^{th}$ subband can be calculated by using the result from [39]. The rate is given by

$$\Gamma_{nm}^{al} = \frac{\sqrt{2m^*}x(1-x)\Omega\Delta U^2}{\hbar^2} D_{nm} \frac{1+2\alpha\varepsilon_f}{\sqrt{\varepsilon_f(1+\alpha\varepsilon_f)}} \quad (8)$$

where $\Omega$ is the volume of the primitive cell, $x$ is the mole fraction, $\Delta U$ is the alloy disorder potential and $D_{nm}$ is overlap integral. Both $D_{nm}$ and $\varepsilon_f$ are defined in Ref. [36]. $\Delta U$ is taken to 0.8 eV for SiGe [40].

## III. RESULTS AND DISCUSSION

The nanowire is taken to be of cross-section 5nm x 5nm. The doping density is taken to be 2 x $10^{25}$/m$^3$. The effective field is taken to be 100 kV/cm. This effective field behaves as a symmetry breaking field and results in Rashba spin orbit coupling. A multisubband Monte Carlo simulation is done. Four subbands[25] are taken into account in each valley. The moderate values of driving electric field allow the majority of electrons to be restricted to the first four subbands. Also the higher subbands will be very higher up in energy owing to small transverse dimensions of the nanowire and thus they can be taken to be depopulated. The energy levels of subbands are determined using an infinite potential well approximation. The material parameters for silicon and germanium are taken to be same as that for bulk silicon and germanium respectively and are adapted from a standard manual on Monte Carlo simulations [30]. The spin-orbit splitting, lande-g factor, lattice constant, permittivity and other parameters for different Ge mole fractions are found out by a linear interpolation [41] between their silicon and germanium values. The bandgap variation with Ge mole fraction, $x$ at 300K and for x<0.85 is given by

$$E_g = 1.12 - 0.41x + 0.008x^2 \text{ eV} \quad (9)$$

The electrons are injected from the source i.e. x=0. A time step of 0.2 fs was selected and the simulation run for 1 x $10^6$ such time steps for the electrons to evolve and reach steady state. Data is recorded for the final 50,000 steps only. The ensemble average is calculated for the components of the spin vector for the last 50,000 steps at each point of the wire.

a) *Dependence of spin dephasing on Ge mole fraction at T=300K and driving electric field of 1kv/cm*

Figure 2 shows the decay of magnitude of ensemble averaged spin vector for different Ge mole fractions of x=0.2, 0.3, 0.5 and 0.8 at room temperature and at a driving electric field of 1kV/cm. The electrons are injected with an initial polarization along the thickness of the wire, i.e. along the z-direction.

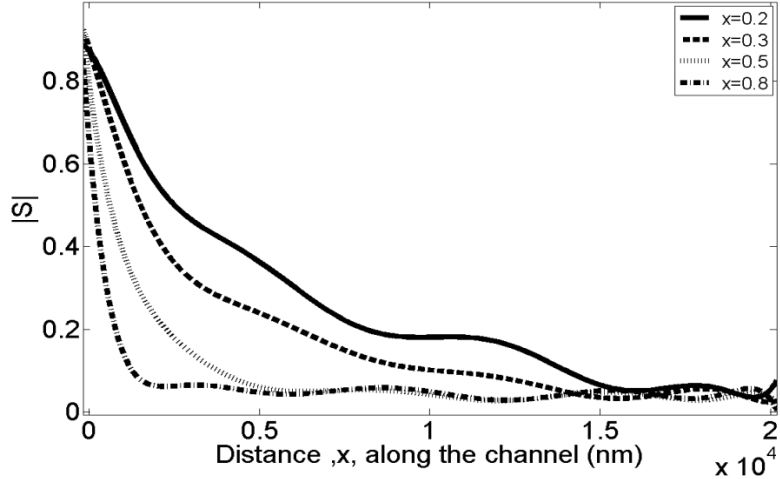

Fig.2 Decay of spin along the channel for $Si_{1-x}Ge_x$ nanowire at different Ge mole fractions (*x*) at 300K, a driving electric field of 1kV/cm with initial injection polarization along the *z*-direction

The spin dephasing length is 4.85 μm at *x*=0.2. It decreases to 2.4 μm at *x*=0.3, to 1.06 μm at *x*=0.5 and to 370 nm at *x*=0.8. Thus the spin dephasing length decreases with the increase in Ge mole fraction in the SiGe nanowire implying faster spin dephasing at higher Ge percentage.

The reduction in spin relaxation length with the increase in germanium mole fraction can be explained as follows on account of three reasons

1. On increasing the germanium content in the alloy, the Rashba spin orbit coefficient $\alpha$ gets stronger, thereby increasing the spin orbit interaction which leads to faster spin dephasing. The variation of the Rashba coefficient $\eta$ with mole fraction at 300K and a effective field of 100kV/cm is shown in Figure 3(a) and has been computed using Eq. (3).
2. On increasing the Ge mole fraction, there occurs an enhancement in the Elliott-Yafet relaxation rate. This enhancement occurs because the spin-orbit splitting increases and the bandgap reduce with the increase in germanium mole fraction. These two combined effects increase the spin flip scattering (in accordance with the Eq. (8)) due to EY mechanism and leads to a faster spin relaxation. The increase of the spin orbit slitting with mole fraction at 300K is shown in Fig. 3(b) and has been computed using a linear interpolation.

3. With increase in the germanium percentage, the dephasing due to scattering increases. This can be seen from Figure 3(c) which shows the average number of time steps of 0.2fs between two scattering events versus mole fraction. The figure depicts that the time interval between two scattering event decreases with increase in Ge mole fraction, thus implying increased scattering. This increase in scattering randomizes the electron spin faster leading to faster dephasing and smaller dephasing lengths.

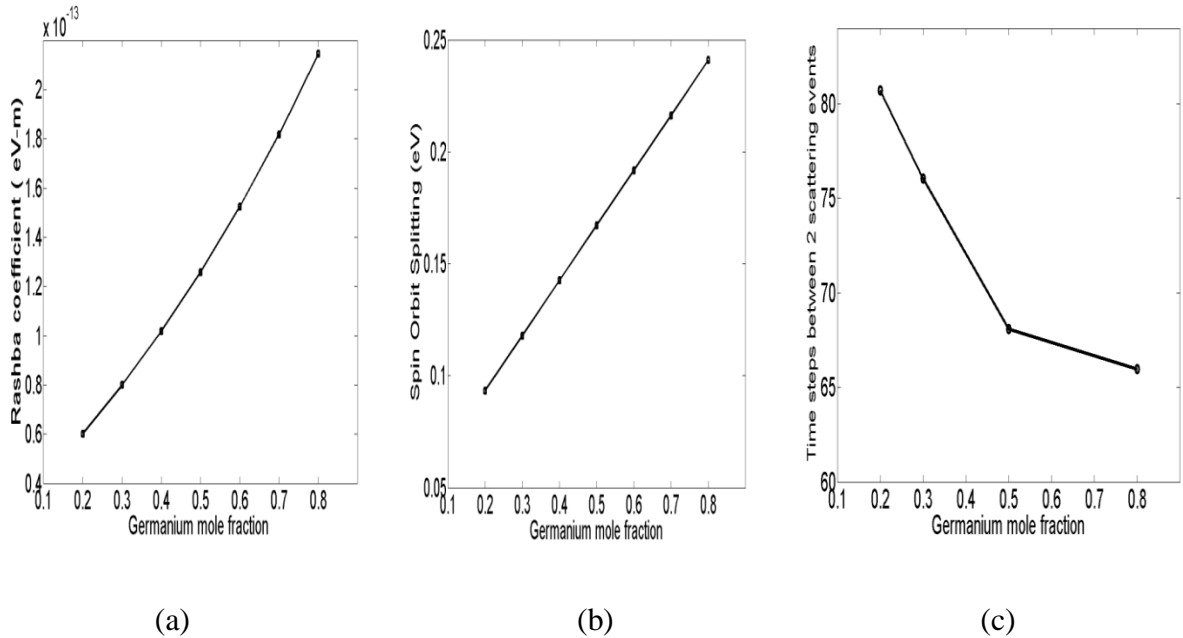

(a) (b) (c)

Fig.3. (a) Rashba coefficient vs. germanium mole fraction at 300K and a effective field of 100kV/cm (b) Spin Orbit Splitting vs germanium mole fraction at 300K (c) Average number of time steps of 0.2 fs between 2 scattering events vs. germanium mole fraction at 300K

b) *Dependence of spin dephasing on Ge mole fraction at T=77K and driving electric field of 1kv/cm*

Figure 4 shows the decay of magnitude of ensemble averaged spin vector for different Ge mole fractions of x=0.2, 0.3, 0.5 and 0.8 at 77K and at a driving electric field of 1kV/cm. The electrons are injected with an initial polarization along the thickness of the wire, i.e. along the z-direction.

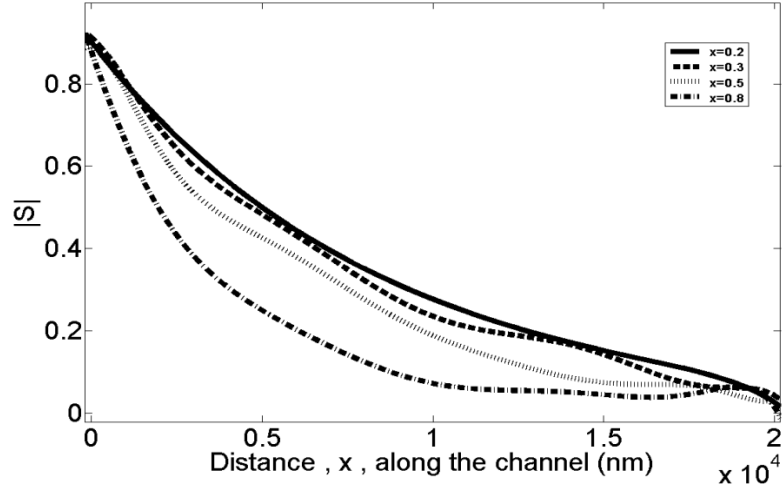

Fig. 4. Decay of spin along the channel for $Si_{1-x}Ge_x$ nanowire at different Ge mole fractions (*x*) at 77K, a driving electric field of 1kV/cm with initial injection polarization along the z-direction

The spin dephasing length is 7.54 μm at *x*=0.2. It decreases to 7.13 μm at *x*=0.3, to 6.19 μm at *x*=0.5 and to 3.70 μm at *x*=0.8. Thus spin dephasing length is a strong function of temperature with the dephasing length showing an increase with the decrease in temperature.

This increase is because of the fact that on decreasing the crystal temperature the phonon scattering rates decrease. The reduced scattering rate thus enables the electron to travel relatively larger distance in between scattering. Thus an electron travels larger distances at lower temperatures before its spin precession vector gets completely randomized due to scattering. Hence the electrons possess larger spin dephasing lengths at lower temperatures.

c) *Decay of spin components and the steady state spin distribution*

i) *X-polarized injection*

Figure 5 shows the decay of the ensemble averaged *x, y* and *z* components of the spin vector along the channel for *x*-polarized injection, i.e. along the length of the channel in the SiGe nanowire for a germanium mole fraction of 0.2. The driving electric field is 1kV/cm and crystal temperature is 300K. Figure 6 shows the steady state distribution of the *x, y* and *z* components of the spin vector of electrons in the ensemble [23] for *x*-polarized injection.

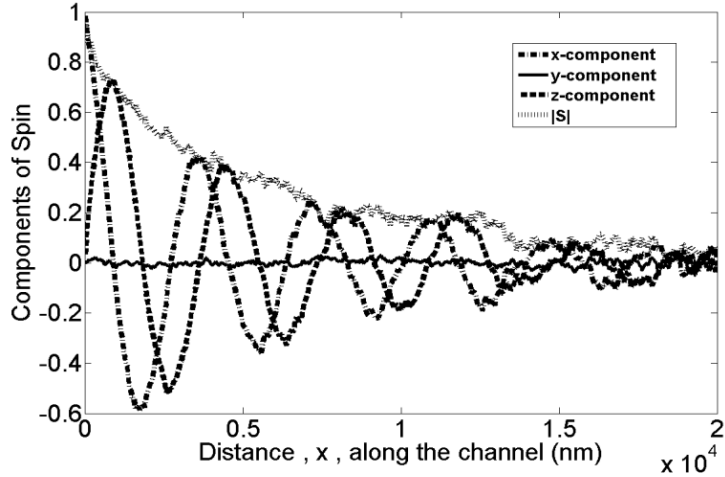

Fig.5. Dephasing of the *x, y* and *z* components of ensemble average spin in $Si_{0.8}Ge_{0.2}$ nanowire at 300K at a driving electric field of 1kV/cm with initial injection polarization along the *x*-direction

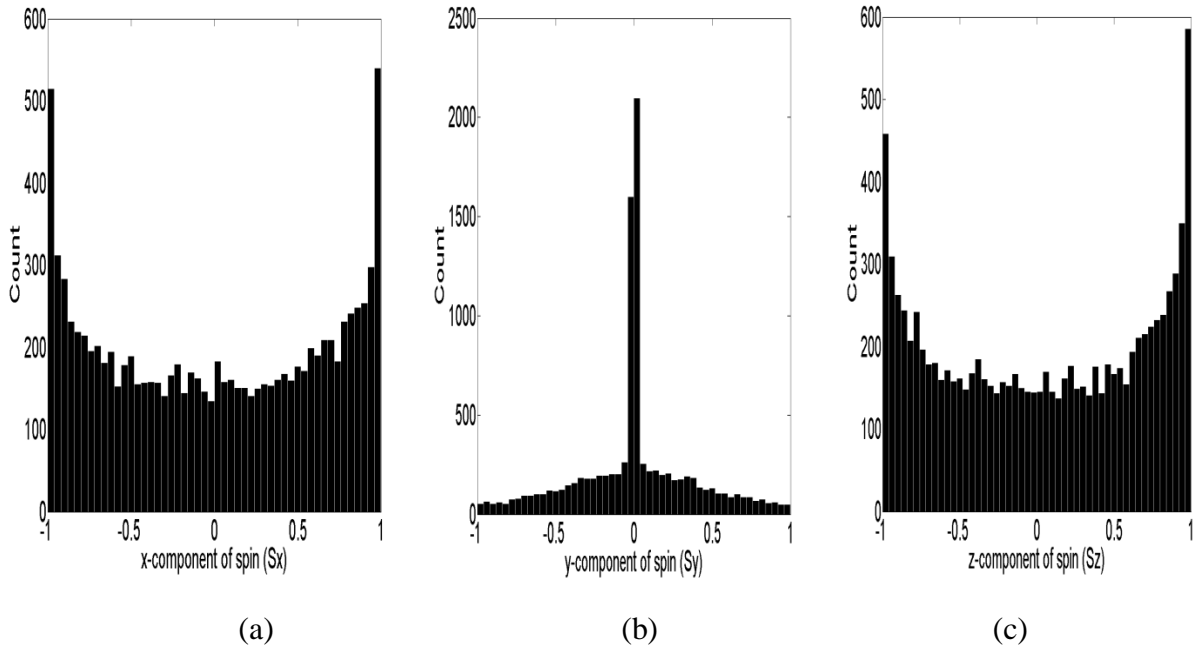

(a)                      (b)                      (c)

Fig.6. Steady state distribution of the spin components in the $Si_{0.8}Ge_{0.2}$ nanowire at 300K at a driving electric field of 1kV/cm with initial injection polarization along the channel length (a)

Distribution of the *x*-component, (b) distribution of the *y*-component and (c) distribution of the *z*-component

Since the initial polarization is in the *x*-direction, i.e. at x=0, $S_x$=1, $S_y$=0, $S_z$=0, therefore the ensemble averaged *y*-component remains equal to zero since the spin orbit interaction does not couple *x*-polarized or *z*-polarized spins to the *y*-polarized spins as seen from the Eq.(5). This can be seen from Fig.5 in which the ensemble averaged *y*-component remains close to zero and from Fig.6 (b) where the *y*-component distribution is like a delta function at zero indicating that most of the electrons have *y*-component very close to or equal to zero.

The *x* and *z* components of the ensemble averaged spin show an oscillatory decay and begin with a π/2 phase shift at *x*=0 which changes due to dephasing. This oscillatory behaviour can be ascertained analytically from the Eq. (4) and Eq. (6). This oscillatory behaviour of the decay of spin components manifests itself in the spin distribution of the spin components which show a U-shaped distribution. More electrons in the ensemble have their spin components closer to +1 and -1 than any other value in between.

   *ii) Y-polarized injection*

Figure 7 shows the decay of the ensemble averaged *x, y* and *z* components of the spin vector along the channel for y-polarized injection in the SiGe nanowire for a germanium mole fraction of 0.2. The driving electric field is 1kV/cm and crystal temperature is 300K. Figure 8 shows the steady state distribution of the *x, y* and *z* components of the spin vector [23] of electrons in the ensemble for y-polarized injection.

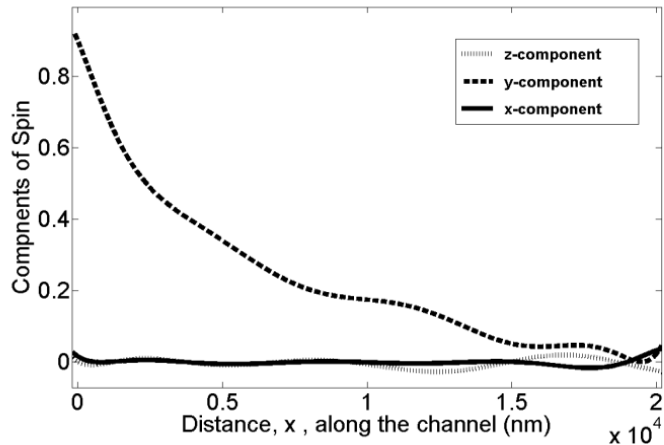

Fig.7. Dephasing of the *x*, *y* and *z* components of ensemble average spin in $Si_{0.8}Ge_{0.2}$ nanowire at 300K, a driving electric field of 1kV/cm with initial injection polarization along the *y*-direction

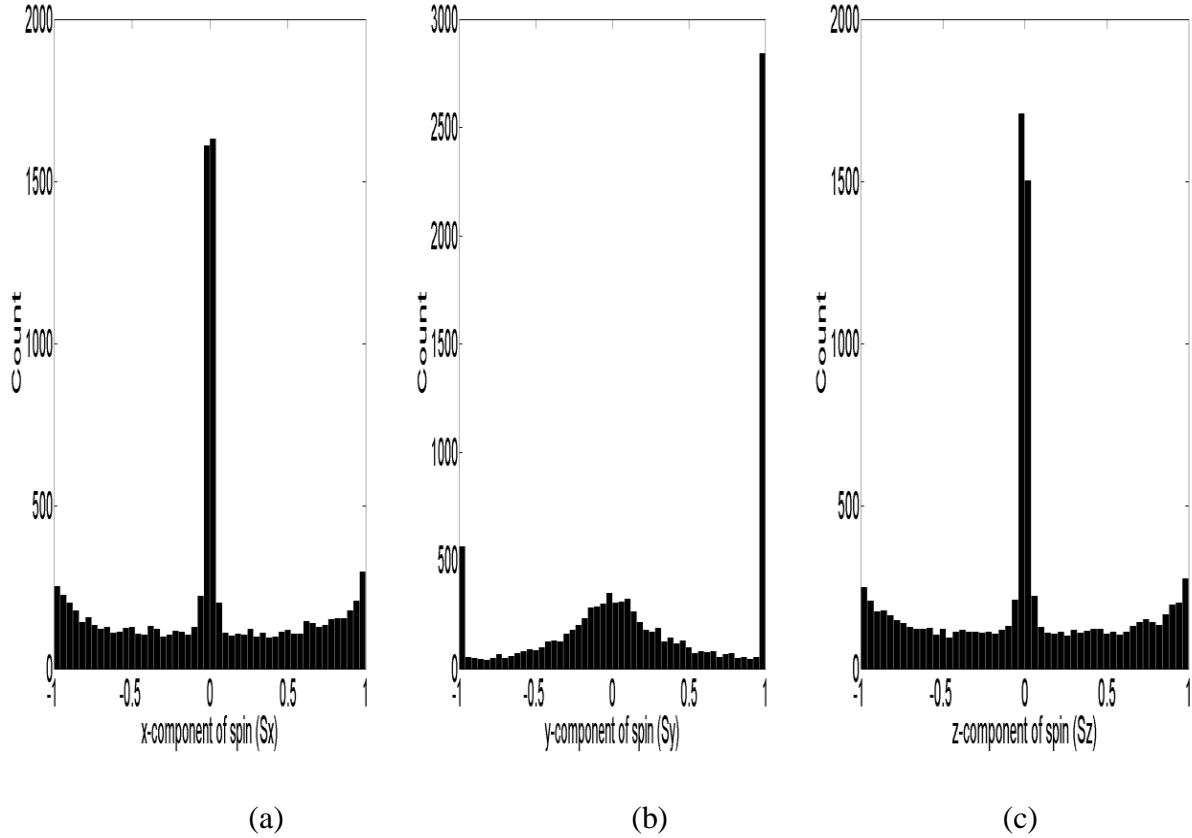

(a)           (b)           (c)

Fig.8. Steady state distribution of the spin components in the $Si_{0.8}Ge_{0.2}$ nanowire at 300K at a driving electric field of 1kV/cm with initial injection polarization along the *y*-direction (a) Distribution of the *x*-component, (b) distribution of the *y*-component and (c) distribution of the *z*-component

Since the initial polarization is in the *y*-direction, i.e. at *x*=0, $S_x$=0, $S_y$=1, $S_z$=0, therefore the injected electrons retain their +1 polarization along the *y*-direction since the spin orbit interaction does not couple *x*-polarized or *z*-polarized spins to the *y*-polarized spins as seen from the Eq.(5). This is evident from Fig.8 (b) where most of the electrons have *y*-component of spin equal to 1. However due to ensemble averaging over all the electrons present in the material, the ensemble averaged *y*-component shows an exponential type decay as seen from Fig. 7.

Since both the *x* and *z* components of spin of the injected electrons are zero to start with they remain at zero since only the *x* and *z* components are coupled to each other but not the *y*-component (in accordance with Eq. (4) and Eq. (6)). This is seen from Fig. 8 (a) and 8 (c). The ensemble averaged *x* and *z* component also remain near zero as in Fig. 7.

### iii) *Z-polarized injection*

Figure 9 shows the decay of the ensemble averaged *x, y* and *z* components of the spin vector along the channel for z-polarized injection i.e. along the thickness of the wire in the SiGe nanowire for a germanium mole fraction of 0.2. The driving electric field is 1kV/cm and crystal temperature is 300K. Figure 10 shows the steady state distribution [23] of the *x, y* and *z* components of the spin vector of electrons in the ensemble for z-polarized injection.

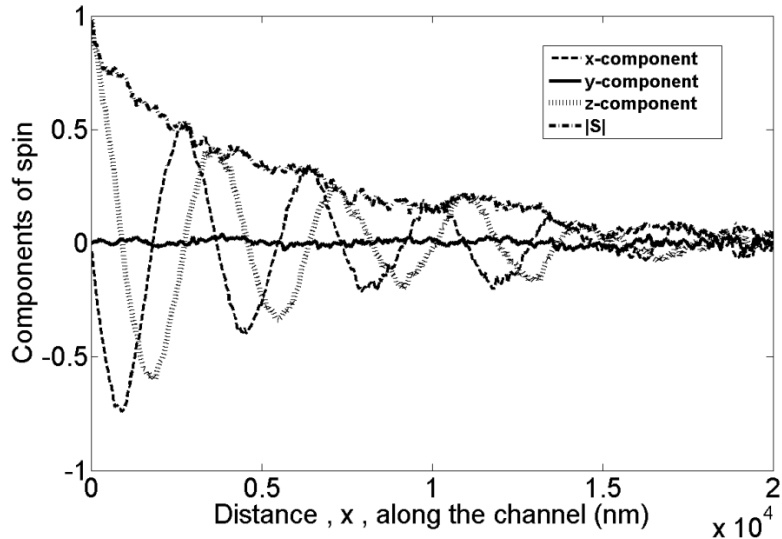

Fig.9. Dephasing of the *x*, *y* and *z* components of ensemble average spin in $Si_{0.8}Ge_{0.2}$ nanowire at 300K at a driving electric field of 1kV/cm with initial injection polarization along the z-direction

Like in the case of *x*-polarized injection, the y-component of the spins of the ensemble remains at zero for *z*-polarized injection as well. This is depicted in Fig.10 (b). Also the ensemble averaged *y*-component remains close to zero as shown in Fig.9.

The *x* and *z* components of the ensemble averaged spin show an oscillatory decay and begin with a π/2 phase shift at *x*=0. However the initial phase for the *x* and *z* components of spin for *z*-polarized injection and for *x*-polarized injection differ from each other. This oscillatory behaviour and the difference in phases can be discerned analytically by solving the Eq. (4) and Eq. (6). This oscillatory behaviour of the decay of spin components manifests itself in the spin distribution of the spin components which show a U-shaped distribution.

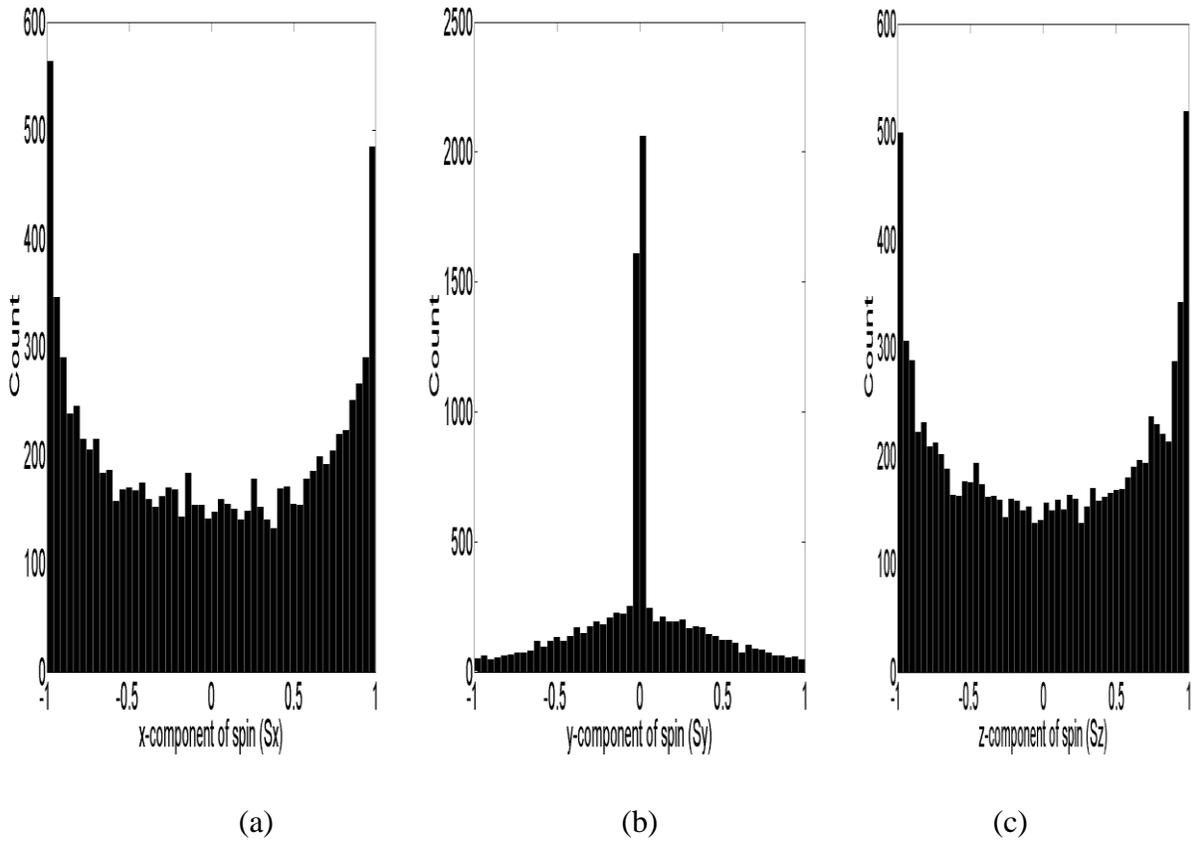

(a) (b) (c)

Fig.10. Steady state distribution of the spin components in the $Si_{0.8}Ge_{0.2}$ nanowire at 300K at a driving electric field of 1kV/cm with initial injection polarization along the z-direction (a) Distribution of the *x*-component, (b) distribution of the *y*-component and (c) distribution of the *z*-component

As has been depicted in this section, spin depolarization is asymmetric in spin orientation [23,24]. The Eq. (4), (5) and (6) bring out this anisotropy in the evolution of spin with time. These equations show the evolution of spin components during each free flight time. During this free flight time there is no dephasing and the evolution of spin is coherent and thus this can be seen as coherent motion or precession of spin. However this is only one side of the picture. The electric field changes the velocity of each particle during the free flight time and thus the effective magnetic field that each particle experiences changes over each free flight time. Also scattering events change the velocities of the electrons randomly at discrete points in time. These two effects cause effective dephasing of spin along the channel when one ensemble averages over the electrons. Thus the entire spin evolution can be considered as coherent motion (precession) and depolarization (decay in magnitude) [23]. There is a continuous competition between the coherent and the incoherent (depolarization) dynamics depending upon the relative magnitudes of spin precession vector and the scattering rates. A dominance of the dephasing rate will lead to spin depolarization and this is manifested in the spin as a monotonic decay of the ensemble averaged spin. On the other hand when the spin precession vector is large, the decay is oscillatory, which is a direct result of coherent dynamics (spin precession).

In silicon germanium (with only Rashba coefficient present), the $x$ and $z$ polarizations of the spin are coupled strongly by the Rashba coefficient as compared to the $y$-polarization which is not coupled to the spin orbit interaction. Therefore the spin precession vector is larger for $x$ and $z$ components of spin and thus these display oscillatory behaviour for $x$- and $z$-polarized injection as already shown. For the $y$-polarized injection, however, the dephasing due to ensemble averaging dominates and hence leads to a monotonic decay.

## IV. CONCLUSIONS

In this paper, we have investigated spin dephasing in a SiGe nanowire. We studied the dependence of spin dephasing on germanium mole fractions in the alloy and found the spin dephasing length to decrease with the increase in germanium percentage. This was attributed to three effects- faster dephasing due to increase in a) Rashba coefficient, b) spin orbit splitting and c) scattering rates. The dephasing lengths increased at lower temperature due to reduction in scattering rates. We have also shown that the steady state distribution of spins in the ensemble and the decay of ensemble averaged spin components depend strongly on the initial direction of spin polarization.